\documentstyle[12pt]{article} 
\begin{document}  
\begin{flushright}   {
DCPT-09-26}\\   
{IPPP-09-13}\\   
SHEP-08-36\\ 
Revised \today 
\end{flushright} 
\vspace*{1.0truecm}  
\begin{center} {\Large \bf \boldmath Explicit CP violation  in the MSSM \\[2mm] 
through $gg\to H_1\to \gamma\gamma$}  

\vspace*{1.0truecm} 
{\large S. Hesselbach$^a$\footnote{stefan.hesselbach@durham.ac.uk}, 
S. Moretti$^b$\footnote{stefano@soton.ac.uk}, 
S. Munir$^c$\footnote{smunir@fisica.unam.mx}, 
P. Poulose$^d$\footnote{poulose@iitg.ernet.in}}  

\vspace*{0.5truecm} 
{\it $^a$ IPPP, University of Durham, Durham, DH1 3LE, UK\\ 
$^b$ School of Physics \& Astronomy,\\ University of Southampton,  Highfield, Southampton SO17 1BJ, UK\\ 
$^c$  Instituto de F\'{i}sica, Departmento de F\'{i}sica Te\'orica,  \\ Universidad Nacional Aut\'onoma de M\'exico, Apartado Postal 20-364,\\   
01000, M\'exico, D.F.\\ 
$^d$ Physics Department, IIT Guwahati, Assam 781039, INDIA} 
\end{center}  
\begin{abstract} 
We prove the strong sensitivity of the $gg\to H_1\rightarrow \gamma\gamma$  cross section at the Large Hadron Collider on the explicitly CP-violating  phases of the Minimal Supersymmetric Standard Model, where $H_1$ is the lightest Supersymmetric Higgs boson.   
\end{abstract}   

\section{Introduction}  
One of the main reasons to build the Large Hadron Collider (LHC) at CERN is the determination of the mechanism of Electro-Weak Symmetry Breaking (EWSB). In the Standard Model (SM) of elementary particle physics and its extensions incorporating Supersymmetry (SUSY), EWSB occurs through the Higgs mechanism, which in turn leads to the existence of one or more Higgs  particles.  

Within the Minimal Supersymmetric Standard Model (MSSM), a realisation of SUSY with minimal particle content and gauge structure, the Higgs potential  conserves Charge \& Parity (CP) at tree level \cite{Higgs-hunter}. Beyond the  trivial order, several studies have shown that CP invariance of the Higgs  potential may in principle be broken by radiative corrections \cite{Maekawa:1992un},  as the Vacuum Expectation Values (VEVs) of the two Higgs doublets can develop a  relative phase \cite{Pilaftsis:1998dd}. This type of CP violation is generally  referred to as spontaneous CP violation and it requires a light Higgs  state as a result of the Georgi-Pais theorem  \cite{Georgi:1974au}, but the possibility of the latter has now essentially been   ruled out by experiment \cite{Pomarol:1992bm}--\cite{Kong:1997ux}.  

CP violation can also be explicitly induced in the MSSM, in much the same way  as it is done in the SM, by complex Yukawa couplings of the Higgs bosons to  sfermions. There are several new parameters  in the SUSY theory, that are absent in the SM, which could well be  complex and thus possess CP-violating phases.    
However,   the CP-violating phases associated with the sfermions of the first and, to a  lesser extent, second generations are severely constrained by bounds on the  Electric Dipole Moments (EDMs) of the electron, neutron and muon. Nonetheless,  there have been several suggestions \cite{Nath:1991dn}--\cite{Cheung:2009fc} to evade these  constraints without suppressing the CP-violating phases.   

By building on the results of  Refs.~\cite{Dedes:1999sj,Dedes:1999zh} (for the production) and \cite{Moretti:2007th}--\cite{Hesselbach:2007gf} (for the decay) -- see also Refs.~\cite{Dedes:2001zf}--\cite{Ellis:2004fs} -- we will look here at the LHC phenomenology of the  $gg\to H_1\rightarrow \gamma\gamma$ process (where $H_1$ labels the lightest  neutral Higgs state of the CP-violating MSSM), which involves the (leading)  direct effects of CP violation through couplings of the $H_i$ ($i=1,2,3$  corresponding to the three neutral Higgs bosons)  to sparticles in the loops as well as the (sub-leading) indirect effects through  scalar-pseudoscalar mixing yielding  the CP-mixed state $H_i$.  (See Ref.~\cite{Hesselbach:2009st} for some preliminary accounts in this respect.) All sources of  CP violation are reviewed in the next Section, where we also lay down our  nomenclature and conventions. In the following Section we present our results,  then we conclude.  

\section{CP violation in the di-photon Higgs search channel}  

Explicit CP violation arises in the Higgs sector of the MSSM when various related  couplings become complex. One consequence is that the physical Higgs bosons  are no more CP eigenstates, but a mixture of these \cite{Pilaftsis:1999qt}--\cite{Frank:2006yh}.  One may then look at the production and decay of the lightest of the physical Higgs  particles, hereafter labeled $H_1$. CP-violating effects in the combined production and decay process enter through complex $H_i$-$\tilde f$-$\tilde f^*$ couplings at production and decay level plus mixing in the propagator  ($\tilde f$ represents a sfermion).  The nature of these couplings are presented in the Appendix.

The leading contribution to Higgs production in gluon fusion is at the one-loop level.  Similarly, Higgs decay into a photon pair also occurs at one-loop level. These loops contain besides SM and charged Higgs states also Supersymmetric particles  ($\tilde{t}_{1,2}$ and $\tilde{b}_{1,2}$ in the case of production while also $\tilde{\tau}_{1,2}$ and $\tilde{\chi}^\pm$ in the case of decay) where the phases of the SUSY parameters enter. Hence CP-conserving and  CP-violating effects enter at {\sl the same perturbative order} in the cross section for  $gg\to {\rm{Higgs}}\to \gamma\gamma$, so that the latter is an ideal laboratory to pursue studies of the complex (or otherwise) nature of the soft SUSY breaking parameters concerned. In contrast, notice that CP-violating effects through mixing in the propagator enter only at higher order, through self-energies, as there is already a tree-level CP-conserving contribution to the propagator\footnote{Also notice that there is also an `indirect' CP-violating contribution to the overall cross section of the process under study if one considers that the $H_1$ (and $H_{2,3}$) mass is subject to similar loop effects, though \cite{Moretti:2007th} has already shown that the consequent effects are marginal (see the left-hand side of Fig. 4 therein), so that  they are included here but not dwelt upon for long.}.   

\begin{figure}[h!] 
\begin{center} 
\vskip 5cm
 \includegraphics{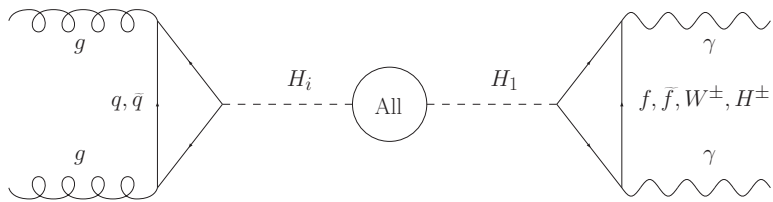}
\end{center} 
\caption{ Leading order Feynman diagram for $gg\rightarrow H_1\rightarrow \gamma\gamma$ including the  effect of mixing in the propagator. } 
\label{fig:feyn} \end{figure}

The dynamics of CP-violating effects in the production and decay stages are rather similar, given that the same diagrammatic topologies are involved (see Fig. \ref{fig:feyn}), in particular, as shown in our previous work \cite{Moretti:2007th}--\cite{Hesselbach:2007gf}, we expect a strong impact of a light stop quark in some regions of the parameter space.  The propagator is considered in the following way. A Higgs particle, $H_i$, produced through gluon fusion,  can be converted into another mass eigenstate, $H_j$, through interaction of  fermion or gauge boson loops and their Supersymmetric counterparts (see Fig. \ref{fig:feyn}).  Therefore,  in the following, when talking about results for the $H_1$, we consider the production of any of $H_i,~i=1,2,3$, which, while propagating, converts into $H_1$. On the one hand, the proper procedure would be to sum over all three Higgs bosons, both in production and decay,  and then integrate over an appropriate $\gamma\gamma$ invariant mass bin, the size of which should be determined by the experimental di-photon mass resolution. On the other hand,  in the regions of parameter space that we will cover  with this study it is rarely the case that two (or more) Higgs boson masses are close together (within the resolution), i.e.,  degenerate, where truly CP violating effects at one-loop level can be  large \cite{Ellis:2004fs}. In fact, in most of the parameter space volume considered, an {\sl isolated} $H_1$ resonance can   be seen by experiment, resolved at the level of about 2 GeV (which is roughly the di-photon mass resolution in ATLAS \cite{ATLAS}, while in CMS it is somewhat  better \cite{CMS}), so that we can treat the decay of the  $H_1$ separately. However, whenever this is not possible, because $M_{H_{2,3}}\le M_{H_1}+2$ GeV, we consider all the degenerate Higgs particles together, as explained in Section 3.   

As intimated, the propagator matrix is obtained from the self-energy of the Higgs particles computed at one-loop level, where we used the expressions provided by \cite{Ellis:2004fs}, which include off-diagonal absorptive parts. The matrix inversion required is done numerically using the Lapack package \cite{lapack}. All relevant couplings and masses are obtained from CPSuperH  version 2 \cite{Lee:2007gn}, which takes into account all applicable experimental  constraints including the low energy EDMs. The cross section   of the full process shown in Fig. \ref{fig:feyn} is computed numerically. The multi-dimensional integration is  carried out using the CUHRE program under the CUBA package \cite{Hahn:2004fe}. For our collider analysis we have used the CTEQ6 Parton Distribution Functions (PDFs) \cite{Pumplin:2002vw}--\cite{Lai:2007dq} computed at the factorisation/renormalisation scale $\mu=\sqrt{\hat{s}}$.  Finally, it should be noted that the results obtained here cannot be simply extracted from existing programs, as these account for CP-violating effects in  production and decay separately, hence without interconnecting them through the appropriately mixed (and off-shell) propagators. Concerning the latter, we are aware that the
inclusion of off-shell Higgs propagators in the diagram of Fig.~\ref{fig:feyn} introduces gauge-violating effects, though
it should be appreciated that these are at most of order $\Gamma_{H_i}/M_{H_i}$, where $\Gamma_{H_i}$ is the
total width of the Higgs state $H_i$ ($i=1...3$). For the mass regimes we are interested in here, particularly for
the case $i=1$, such gauge-violating effects are negligible compared to the typical CP-violating ones that we
will illustrate in the remainder of the paper.  

\section{Results}  

In order to illustrate the typical effects of CP violation in the MSSM,  we have considered a few sample parameter space points and studied the effect of, in particular, light sparticles in the loops, chiefly, of top squarks.  We fix the following MSSM parameters which play only a minor role in CP violation studies for Higgs production and decay:\\  

\(M_1=100~{\rm GeV},~~~M_2=M_3=1~{\rm TeV},~~~M_{Q_3}=M_{D_3}=M_{L_3}=M_{E_3}=M_{\rm SUSY}= 1~{\rm TeV}.\)\\  

We consider the case of all the third generation trilinear couplings being  unified into one single quantity, $A_f$. All the soft masses are taken to be the same at some unification scale, whose representative value adopted here is 1 TeV. When considering the light stop case we take a comparatively light value for $M_{U_3}\sim 250$ GeV, which corresponds to a stop mass of around 200 GeV,  otherwise $M_{U_3}$ is set to 1 TeV.  We could, alternatively, consider small values for $M_{Q_3}$ to reach light squarks, but the effects would qualitatively be the same. So, in   the following we keep a fixed value of $M_{Q_3}=1 $ TeV. In the Higgs scalar-pseudoscalar mixing the product of $\mu A_f$ is relevant rather than $\mu$ or $A_f$ separately. As argued in our earlier works, the only phase that is relevant is thus the sum of the phases of $\mu$ and $A_f$. In our analysis we have kept $\phi_{A_f}=0$ and studied the effect of CP violation by varying $\phi_\mu$. Regarding the absolute 
 values of $\mu$ or $A_f$, in our numerical analysis, we have varied these parameters between 1 and 1.5 TeV. $M_{H^+}$ is instead varied between 100 and 300 GeV. The mass of the lightest Higgs particle is consequently in the range of 50--130 GeV.  We then analyse cases with different $\tan\beta$'s. In particular, low $\tan\beta$ values give very small deviations from the corresponding  CP-conserving cases,  while large $\tan\beta$ values produce significant differences. Also, we take a  representative value of $\tan\beta=20$ to see the effect of the other parameters. In addition,  to illustrate the cases of very large $\tan\beta$ points, we present the  analysis with $\tan\beta=50$ whilst keeping all other parameters constant.  

The main aim of this article is to understand the effect of sparticle  couplings in  the CP-violating MSSM.  As mentioned earlier, effects of CP violation enter  the process considered and combine in three different ways, viz. (i) physical Higgs particles are a mixture of the CP eigenstates, (ii) through the sparticle-Higgs couplings in the production and decay  amplitudes, (iii) through the Higgs mixing in the propagator. Notice though that (i) (and also (iii) to some extent) is a result of (ii), however, in our specific process, (i)--(iii) all enter giving (quantitatively and qualitatively) different effects, so we have listed them separately. As explained in the introduction,  effects of (i) and (ii) in the case of $gg\to H_1$ production and  $H_1\rightarrow \gamma\gamma$ decay  were already reported on in detail in previous literature, albeit separately, so our intent here is to combine them. Furthermore,  we would like to explore the effects due to (iii) in our specific
   process following the general analysis of Ref.~\cite{Ellis:2004fs}.   
\begin{figure}[h] 
\vskip 8cm
 \includegraphics{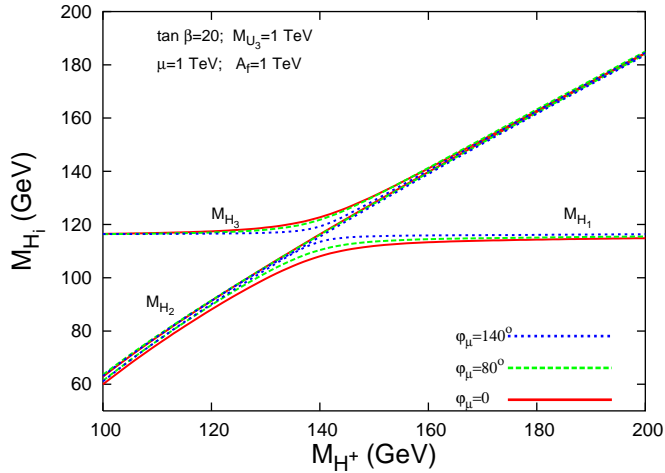} 
\caption{Masses of the physical Higgs bosons against the input parameter $M_{H^+}$ for $\tan\beta=20$, $A_f=1$ TeV, $\mu=1$ TeV and $M_{U_3}=1$ TeV.}  
\label{fig:omix-hmass} 
\end{figure}

In Fig. \ref{fig:omix-hmass} we plot the physical Higgs mass $M_{H_1}$ against $M_{H^+}$ for a typical parameter space point with $\mu=1$ TeV, $A_f=1.5$ TeV and $\tan\beta=20$,  considering three different values of $\phi_\mu=0,~~80^{\rm o},~~140^{\rm o}$. Here, one can appreciate that the   CP-conserving case of  $\phi_\mu=0$ does not present any $M_{H_1}$ point significantly degenerate with either $M_{H_2}$ or $M_{H_3}$  for the entire range of $100\leq M_{H^+}\le 300$ GeV considered, as the mass difference $M_{H_{2,3}}-M_{H_1}$ is always larger than 2 GeV. In fact, in all the cases we considered,  in the CP-conserving case $H_1$, which is the lightest of the physical Higgs particles, is a pseudoscalar at low $M_{H^+}$ values, which turns to  a scalar at large $M_{H^+}$ values. The transition happens around $M_{H^+}=140 - 150$ GeV, where it is possible for both the physical scalar particles to have masses such that $M_{H_2}-M_{H_1}\le 2$ GeV. For the particular $\mu$, $A_f$ and $\tan\beta$ values considered in Fig.~\ref{fig:omix-hmass}, this is not the case. Away from this ``mixing region", the two scalars are well separated in mass. When CP is violated, scalar-pseudoscalar mixing is possible for a large range of $M_{H^+}$ values as  exemplified in   
Tab.~\ref{table:mh}, where $M_{H^+}$ and $M_{H_i},~~i=1,2,3$, are tabulated.   This is done for the same set of parameter values as in the case of Fig.~\ref{fig:omix-hmass}.   

\begin{table} 
\begin{center} 
\begin{tabular}[ht]{|c||r|r|r||r|r|r|} 
\hline 
&  \multicolumn{3}{c||}{$\phi_\mu=80^{\rm o}$}& \multicolumn{3}{c|}{$\phi_\mu=140^{\rm o}$}\\[2mm]\cline{2-7} 
$M_{H^+}$ & $M_{H_1}$& $M_{H_2}$& $M_{H_3}$& $M_{H_1}$& $M_{H_2}$& $M_{H_3}$\\[2mm] \cline{1-7} 
&&&&&&\\  
103. &64.4    &66.3    &120.4   &63.9  &  65.7 &120.4\\  
110. &75.0    &76.6    &120.5   &74.6  &  76.1 &120.4\\  
115. &82.0    &83.5    &120.6   &81.8  &  83.1 &120.5\\  
120. &88.8    &90.2    &120.7   &88.6  &  89.8 &120.5\\  
125. &95.2    &96.6    &121.0   &95.2  &  96.3 &120.6\\  
130. &101.4   &102.9   &121.3   &101.6 & 102.6 &120.8\\  
135. &107.1   &109.0   &122.0   &107.8 & 108.7 &121.1\\  
&&&&&&\\ \hline \end{tabular} 
\caption{Masses of the physical Higgs bosons along with the input parameter $M_{H^+}$ for $\tan\beta=20$, $A_f=1.5$ TeV, $\mu=1$ TeV and  $M_{U_3}=1$ TeV in the degenerate case, where  $M_{H_{2,3}}\leq M_{H_1}+2$ GeV. All masses are in GeV.}  
\label{table:mh} \end{center} \end{table}  

\begin{figure}[h] 
\vskip 8cm 
\includegraphics{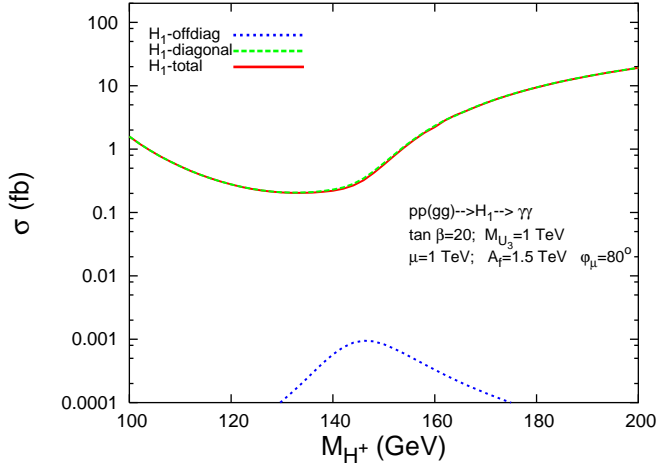} 
\caption{Different contributions of the propagator matrix to the process $gg\rightarrow H_1\rightarrow \gamma\gamma$. The parameters considered are $\tan\beta=20$, $A_f=1.5$ TeV, $\mu=1$ TeV and $M_{U_3}=1$ TeV. Here, the green (dashed) line is the  contribution which comes from the (diagonal) $H_1-H_1$ element in the propagator matrix  and the blue (dotted) line comes from the (off-diagonal) $H_1-H_2$ and $H_1-H_3$ elements, the red (solid) one being their sum.}  
\label{fig:prop} \end{figure}

Fig. \ref{fig:prop} shows the typical effects of CP violation on the cross section through the propagator.  In order to understand the latter, we analyse the $H_1$ production, propagation and decay process in the following way. As intimated already, when the propagator is computed at one-loop level, the considered process could go through a production of any of the three Higgs flavours eventually converting into the $H_1$ state. To understand the effect of Higgs mixing in the propagator, we plot the contribution due to the diagonal elements in the propagator matrix and the contribution due to the off-diagonal elements  separately, along with the total contribution.  The off-diagonal CP-violating terms contribute around 1\% near the degenerate mass region ($M_{H^\pm}\approx 145$ GeV) and much less when the Higgs masses are distinguishably different. This effect is seen in all the parameter space points considered in fact, therefore, it is a general feature of the process.  

\begin{figure}[h] \vskip 8cm 
\includegraphics{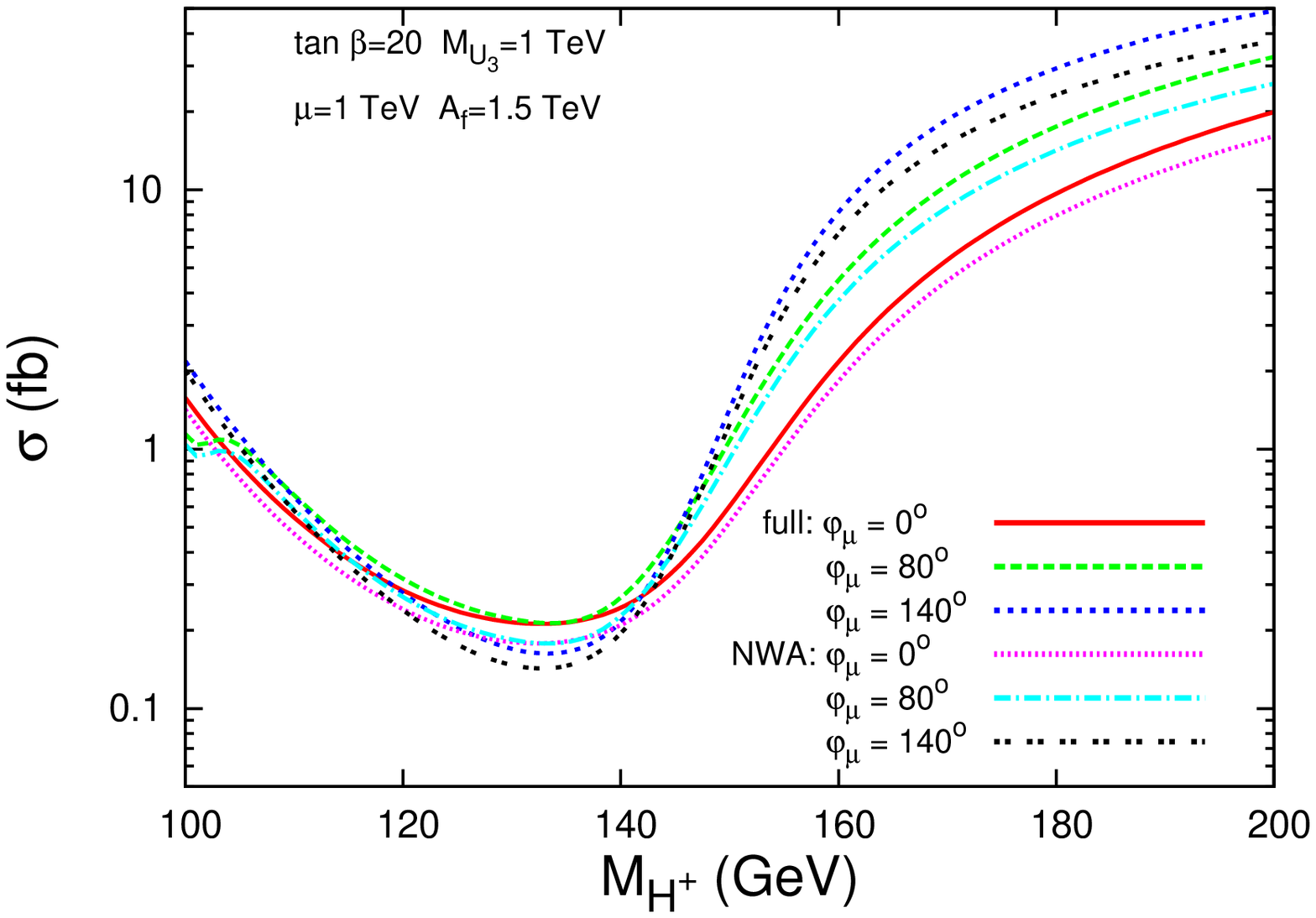} 
\includegraphics{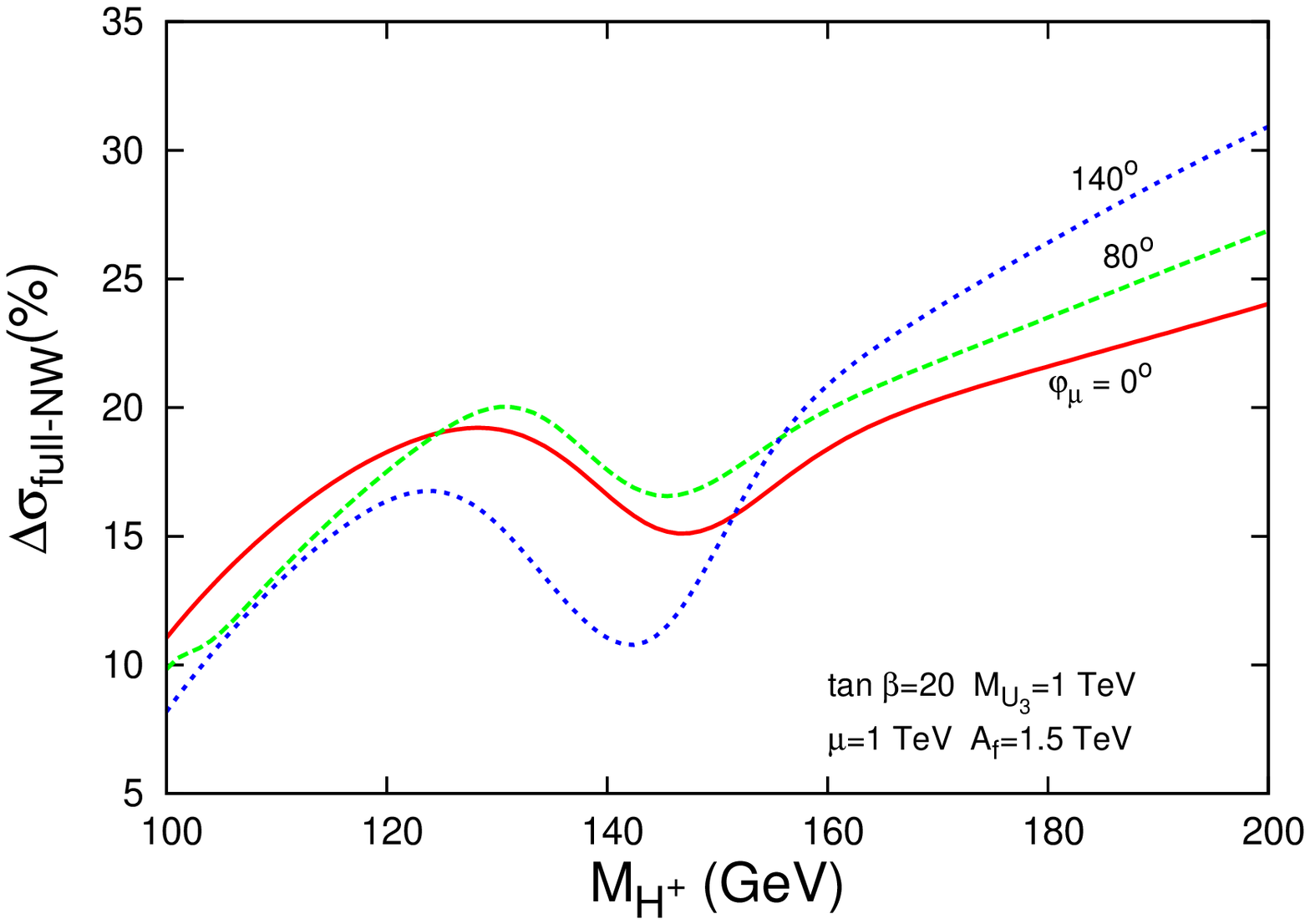} 
\caption{{\em Left:} Cross section of  $gg\rightarrow H_1\rightarrow \gamma\gamma$ with NWA and the full process (with $H_2$ added whenever $M_{H_2} \le M_{H_1}+2$ GeV) for three different values of $\phi_\mu$ plotted as function of $M_{H^+}$.  {\em Right:} Percentage deviations of full cross section from that obtained with NWA plotted against $M_{H^+}$.  The other parameters considered are $\tan\beta=20$, $A_f=1.5$ TeV, $\mu=1$ TeV and $M_{U_3}=1$ TeV.}  
\label{fig:BW} \end{figure}

In the degenerate region, where two Higgs bosons are indistinguishable, their contributions to the total cross section will have to be  added at the amplitude level, and the effect is expected to be distinct from the case of Narrow Width Approximation (NWA).  In Fig.~\ref{fig:BW} (left) we plot the cross section of our process for the two cases in the three situations with $\phi_\mu=0$, $\phi_\mu=80^{\rm o}$  and $\phi_\mu=140^{\rm o}$, keeping all other parameters fixed. In this particular case, $H_1$ and $H_2$ are degenerate at low $M_{H^+}$ values when $\phi_\mu=80^{\rm o}$ and $\phi_\mu=140^{\rm o}$.   
In Fig.~\ref{fig:BW} (right) we plot the percentage deviation of cross section from its value obtained in the NWA, defined as 
\begin{equation} \Delta\sigma(\%)=\frac{\sigma_{full}-\sigma_{NWA}}{\sigma_{full}}\times 100. \nonumber \end{equation} 
The deviation is between 10 to 30\% for $100~~ {\rm GeV}\leq M_{H^+}\leq  200~~ {\rm GeV}$, thereby confirming that a production times decay calculation in NWA, neglecting amplitude level superposition in the degenerate case, 
would lead here to somewhat incorrect results.  
Such large difference is not expected to arise from higher order corrections.  

In summary, we would like to point out that it was {\it a priori} not obvious what the effect of propagator mixing would have been, how CP-violating effects in production and decay canceled or supported each other and how the proper accounting for the degenerate mass cases would have impacted the final results. Our analysis makes it clear though that, while the propagator mixing does not affect the considered process in a significant way, the other two work in tandem to make the overall effect significant.  
While this statement has been illustrated for some  sample parameter space points, we have verified that the above conclusions remain the same for parameter regions with large values of $\tan\beta$, $\mu$ and $A_f$, where the CP-violating effects are prominent. Finally, as Fig.~\ref{fig:BW} incorporates CP-violating effects due to the change in the Higgs couplings with Supersymmetric particles, it is clear that it is the latter which accounts for essentially all the differences between the CP-conserving and CP-violating results, throughout. In this connection, it is of particular importance to investigate the effect of the (dominant)  Higgs-stop-stop couplings with varying stop mass,  as illustrated in the discussion below.   

\begin{figure}[h] \vskip 8cm \includegraphics{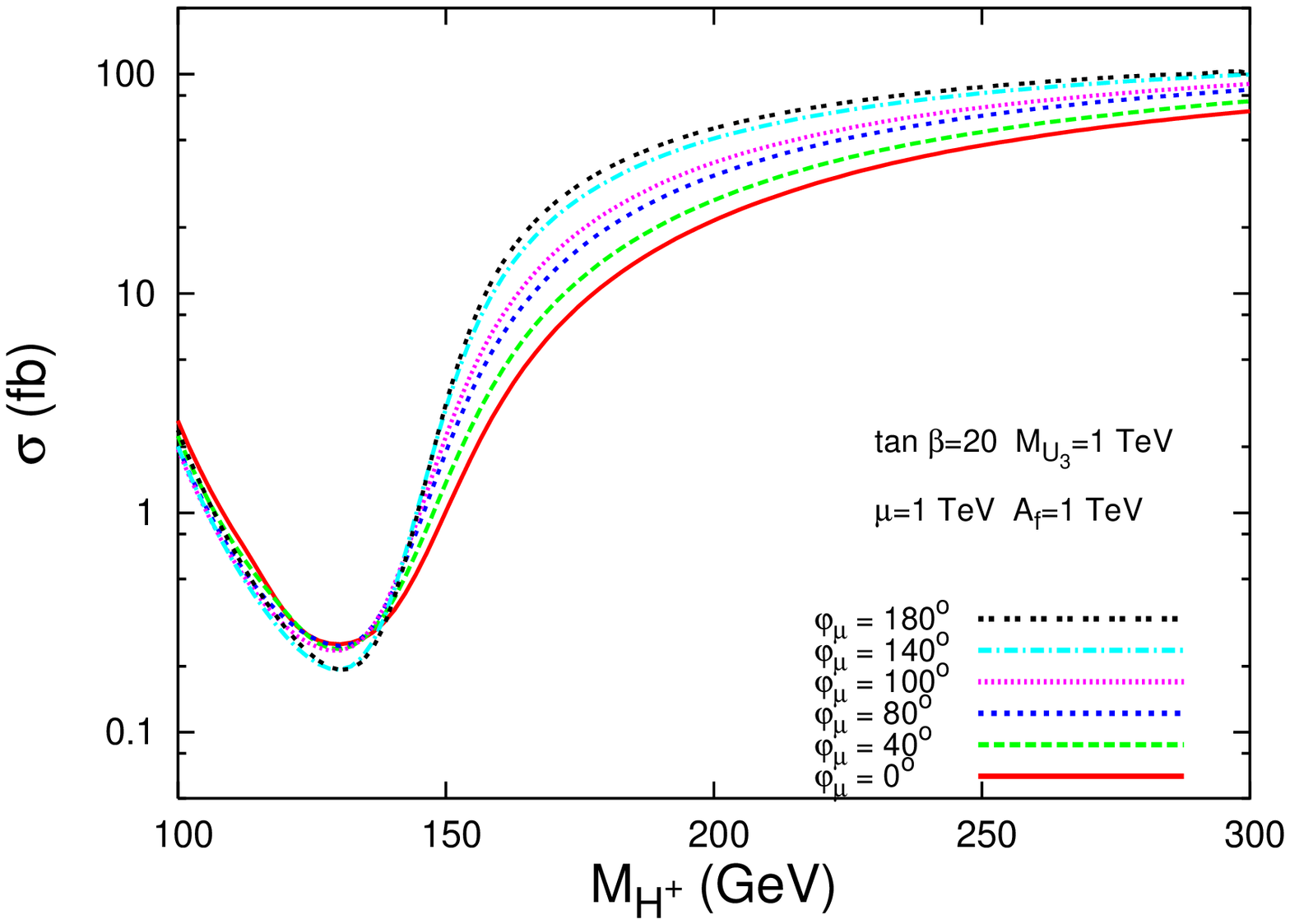} \includegraphics{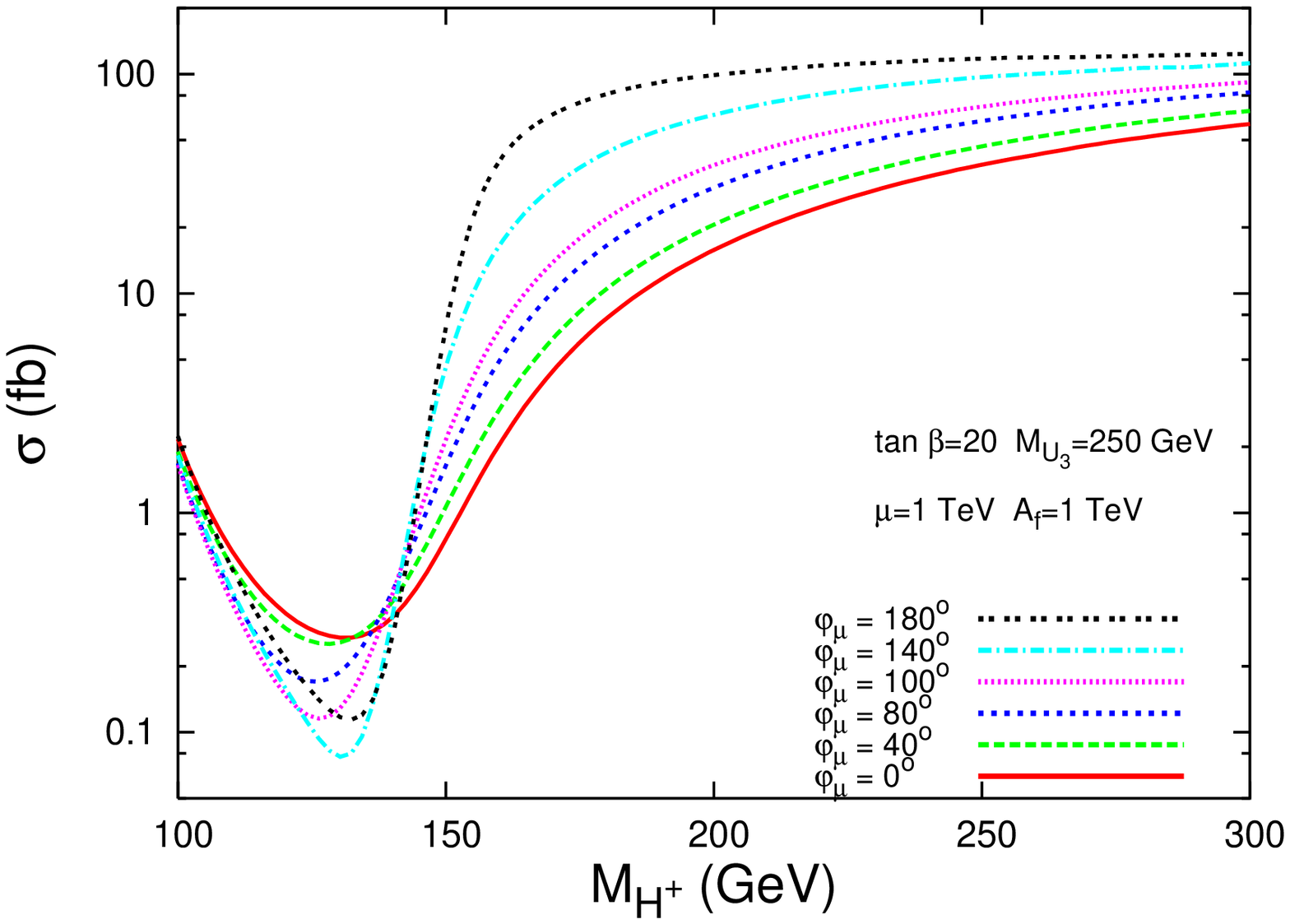} 
\caption{The cross section $\sigma (pp\rightarrow H_1 \rightarrow \gamma\gamma)$ at the LHC plotted against the charged Higgs mass for different $\phi_\mu$ values. The relevant MSSM variables are set as follows: $\tan\beta=20$, $A_f=1$ TeV, $\mu=1$ TeV, with $M_{U_3}=1$ TeV (left plot) and $M_{U_3}=250$ GeV (right plot).}  
\label{fig:tb20A1mu1phimu} \end{figure}

In Fig. \ref{fig:tb20A1mu1phimu}  we plot the full cross section for $gg\to H_1\to \gamma\gamma$  against $M_{H^+}$\footnote{Here and in the rest of the article, $H_1$ means the lightest physical Higgs boson if it is experimentally distinguishable from the other two Higgs bosons. Whenever the lightest one is instead degenerate with one or both of the other two Higgs bosons (in fact, only with $H_2$ for our purposes), effects of all the degenerate states are considered at the amplitude level.}. We have considered $\mu=1$~TeV, $A_f=1$~TeV and $\tan\beta=20$.  There is appreciable variation of the cross section with  $\phi_\mu$. Comparing the two cases of light and heavy stops, it is clear that the effect of the Higgs-stop-stop coupling is significant. Indeed, this was also noticed when we studied the di-photon decay \cite{Hesselbach:2007jf}.   

\begin{figure}[h] \vskip 8cm \includegraphics{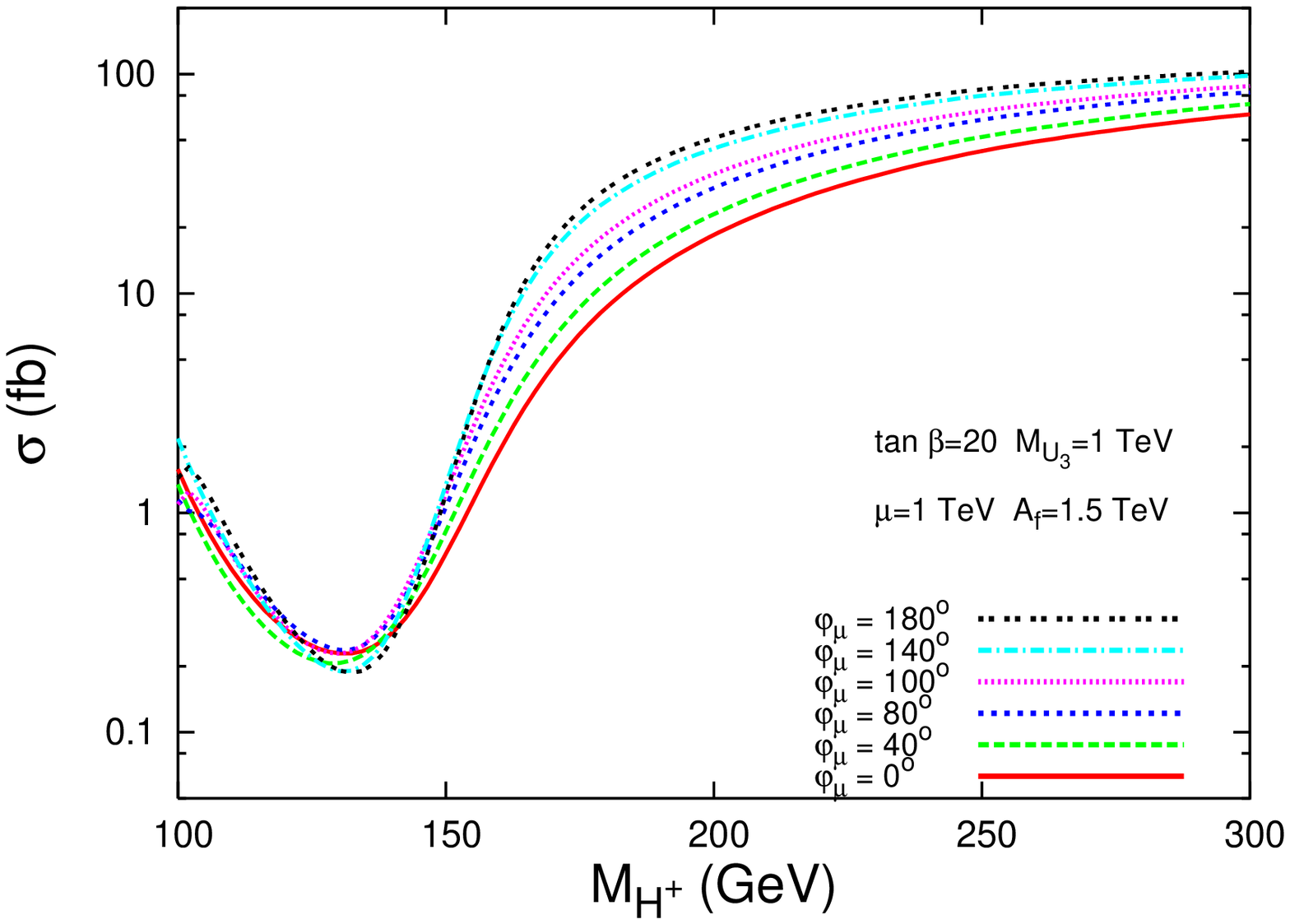} \includegraphics{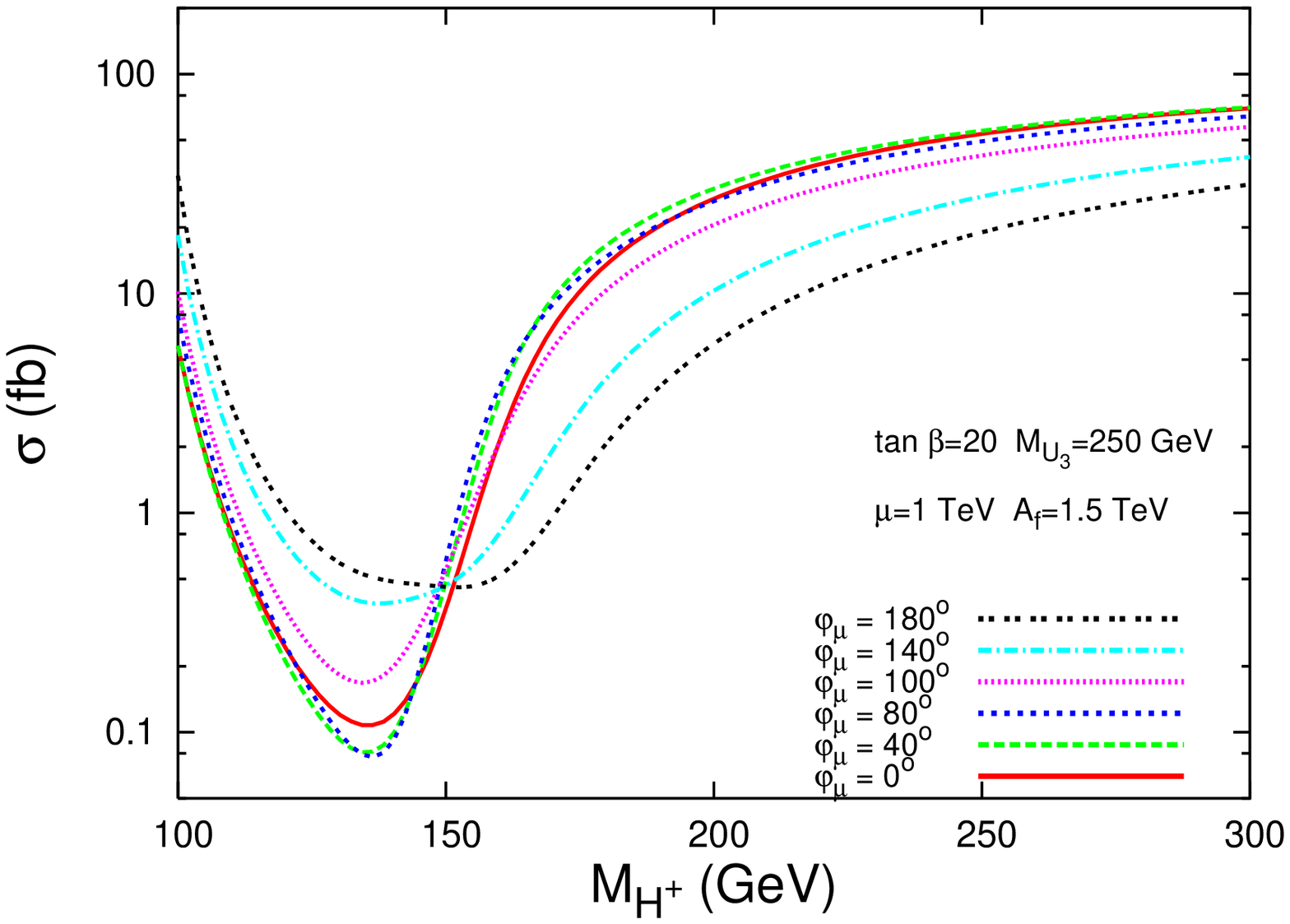} 
\caption{Same as Fig. \ref{fig:tb20A1mu1phimu}, but with $A_f=1.5$ TeV } 
\label{fig:tb20A1.5mu1phimu} \end{figure}

\begin{figure}[h] \vskip 8cm \includegraphics{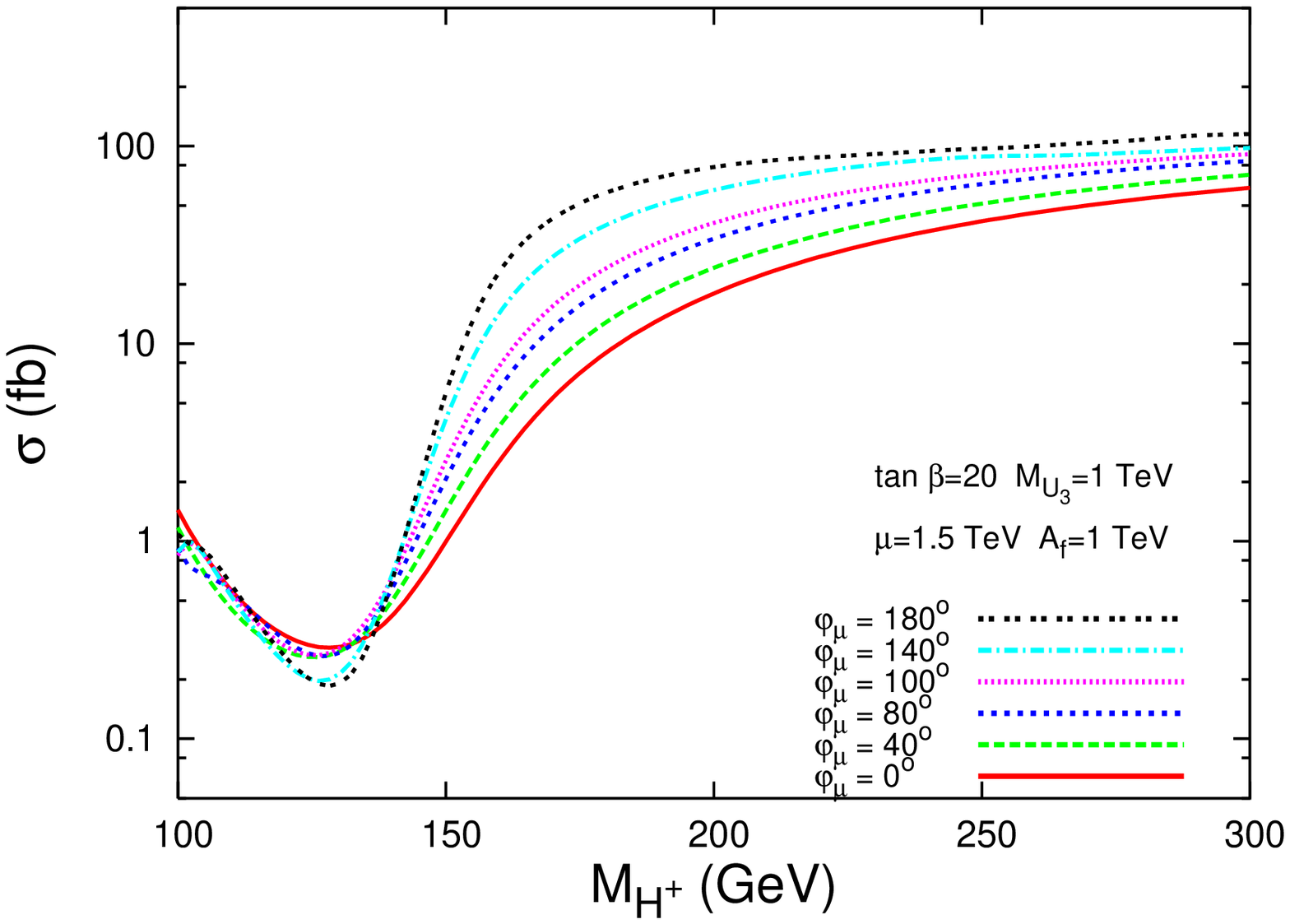} \includegraphics{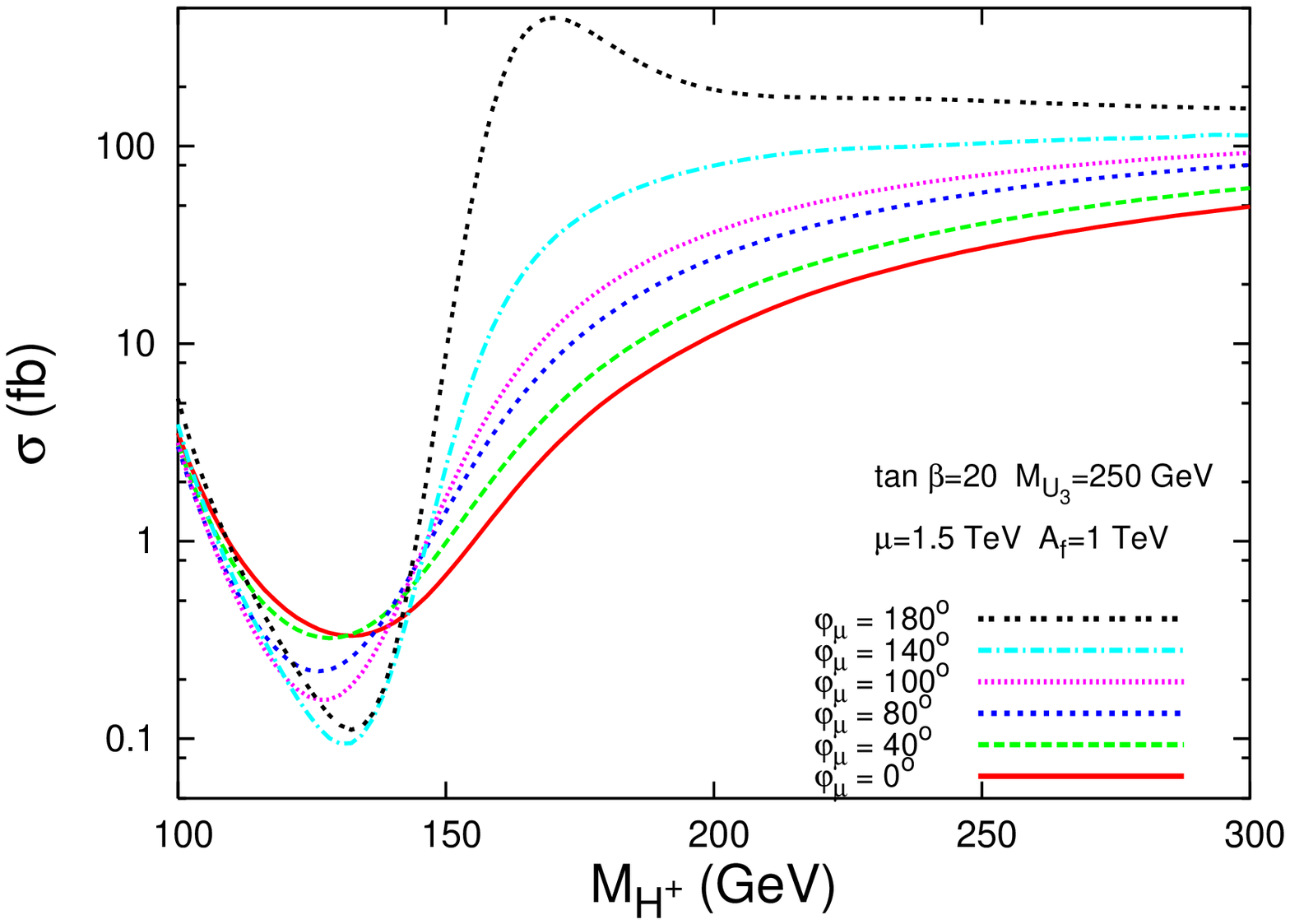} 
\caption{Same as Fig. \ref{fig:tb20A1mu1phimu}, but with $\mu=1.5$ TeV} 
\label{fig:tb20A1mu1.5phimu} \end{figure}

Figs. \ref{fig:tb20A1.5mu1phimu} and \ref{fig:tb20A1mu1.5phimu} illustrate similar studies with $A_f=1.5 ~{\rm TeV}$, $\mu=1$ ${\rm TeV}$ and  $A_f=1$ TeV, $\mu=1.5$ TeV, respectively, in particular  showing how significantly the result depends on $A_f$ and $\mu$. Unlike the case of Fig.  \ref{fig:tb20A1mu1phimu}, where there is only a quantitative difference  between the two cases of light and heavy stop, in  Fig. \ref{fig:tb20A1.5mu1phimu} we see that there is also a  qualitative difference between the two cases (i.e., in the shape of the cross section curve).  In contrast, Fig.~\ref{fig:tb20A1mu1.5phimu} does only show  a quantitative change, yet this simply has to do with  the particular values of $\mu$ and $A_f$ considered. In general, we may  expect the difference between the CP-conserving and CP-violating cases to depend quite sensitively on $A_f$ and $\mu$.  

\begin{figure}[h] \vskip 8cm \includegraphics{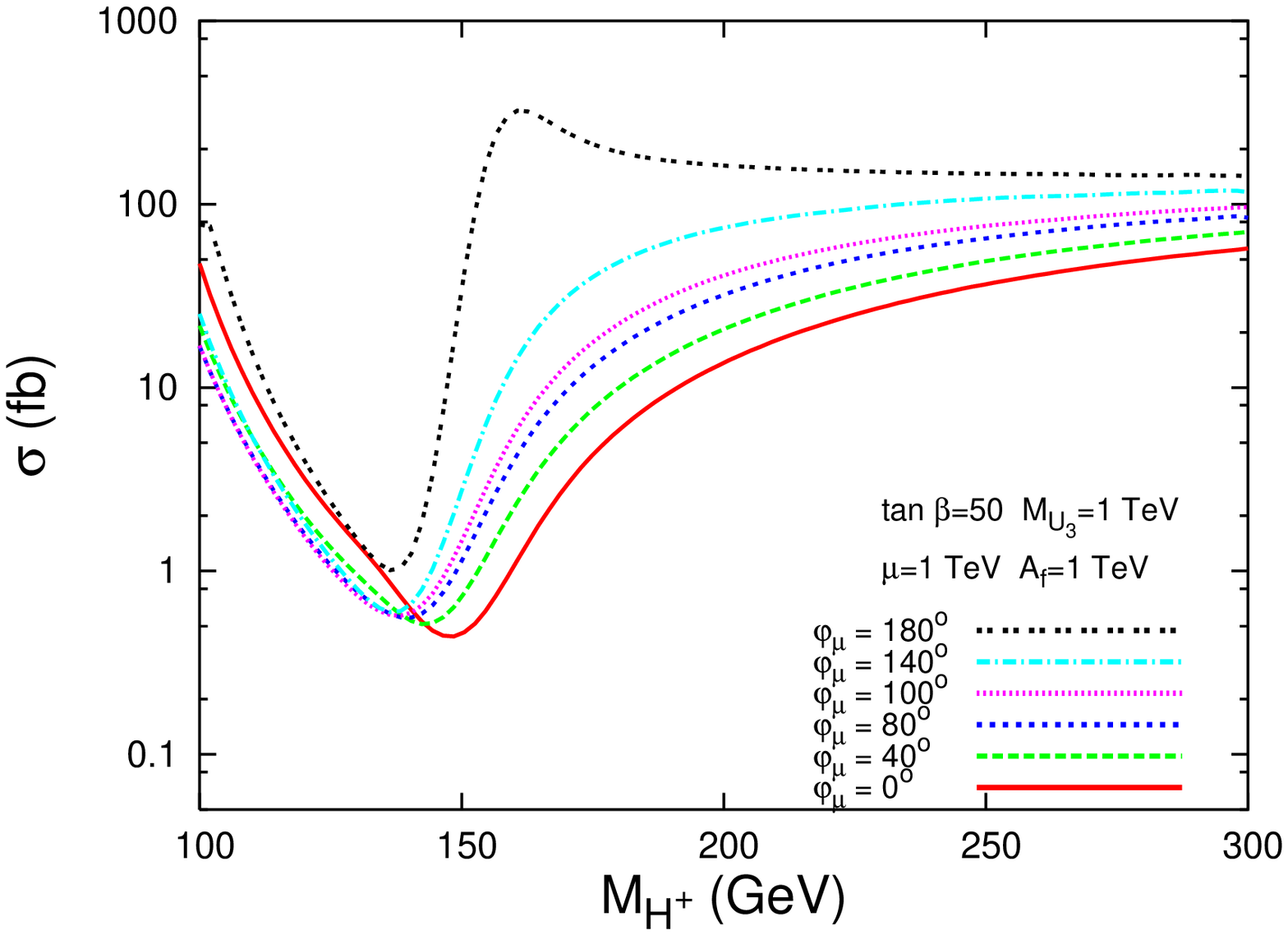} \includegraphics{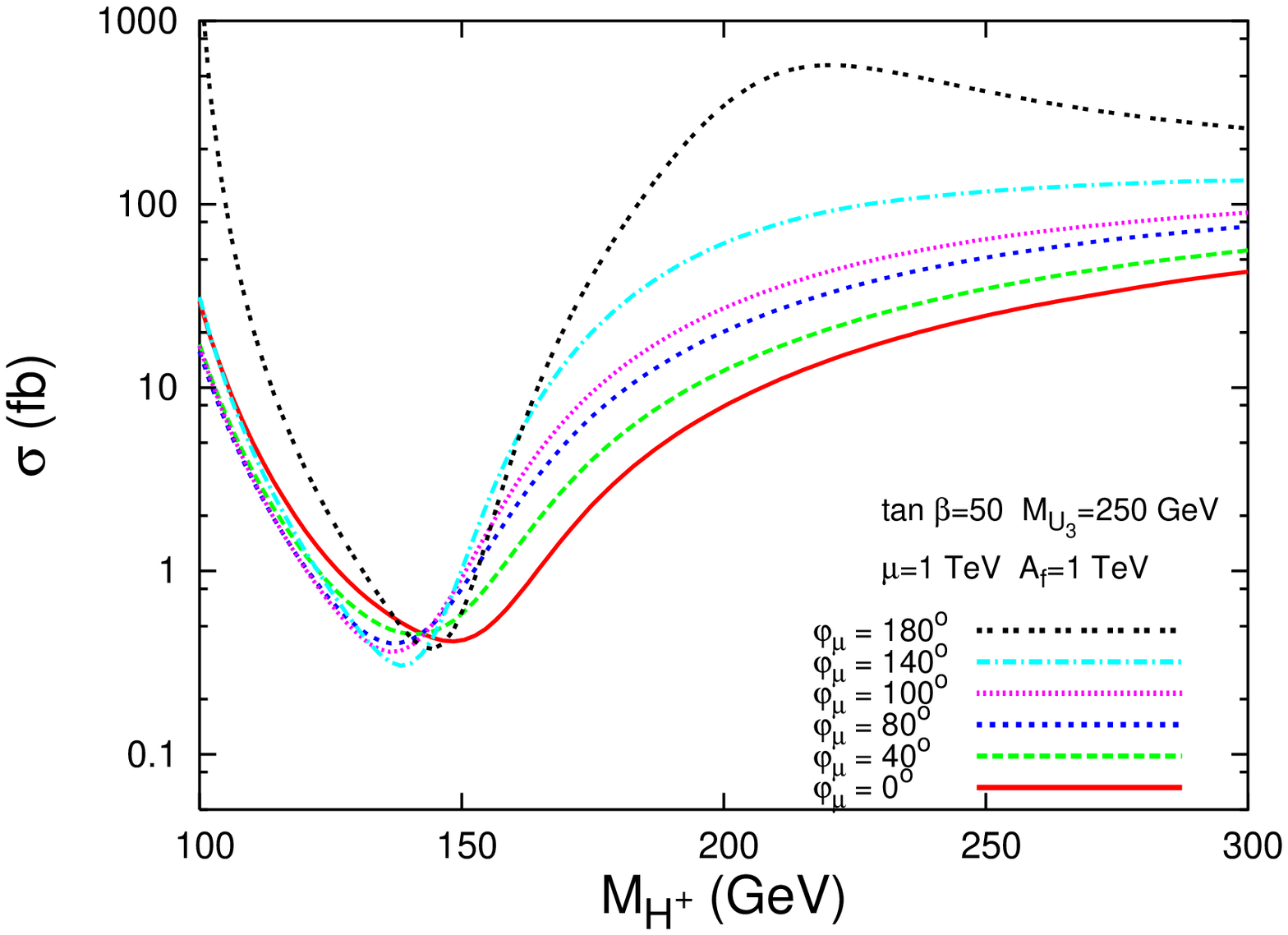} \caption{Same as Fig. \ref{fig:tb20A1mu1phimu}, but with $\tan\beta=50$.} \label{fig:tb50A1mu1phimu} \end{figure}

Fig.~\ref{fig:tb50A1mu1phimu} shows the cross  section for $\tan\beta=$ 50 with $A_f=1$ TeV  and $\mu=1$ TeV.  As in the case of a di-photon decay  considered independent of the production, here too (i.e., for the full process) cases with small $\tan\beta (= 5)$ with $A_f=1$ TeV and $\mu=1$ TeV are not very sensitive to CP-violating phases. 

We may note that the couplings $H_i\tilde t_j\tilde t_k$, Eq. (\ref{eqn:Hisfsf}), as well as  other Higgs couplings, have complicated relations with various SUSY parameters, and therefore it is difficult to understand all the features of the cross section in full detail. Nonetheless we have attempted doing so. We can certainly say that the minimum in the cross section plots  (Figs. 4--8) corresponds to the cross-over region in Fig. 2. Besides,  in all plots presented, the cross section rises (the more the larger $M_{H^\pm}$, so long that it is above 145 GeV or so) with $\phi_\mu$ except in the right plot of Fig. 6 (similarly to the top-right frame of Fig. 11 in \cite{Hesselbach:2007en}) and this can be traced back to the shape of the $H_i\tilde t_j\tilde t_k$ coupling itself. In fact, we generally recover in the case of the full process $gg\to H_1\to\gamma\gamma$ the trends already seen in Ref. \cite{Hesselbach:2007en} for the di-photon decay alone, signifying that CP-violating effects at production level are not offsetting those affecting the decay (as those induced by propagation are negligible), in fact  both are at times cooperating to enhance the overall corrections. Also, one can often appreciate sudden variations in the di-photon yield due to indirect width effects.
For example, the small hump observed in Fig.~\ref{fig:tb20A1mu1.5phimu} 
and in Fig.~\ref{fig:tb50A1mu1phimu} for $\phi_\mu=180^o$ around 
$M_{H^+}=160$ GeV is an indirect effect on the $BR(H_1\rightarrow 
\gamma\gamma)$ due to a dip in the total width of $H_1$ corresponding 
to these parameter values. The dip in the total width itself arises 
because $H_1$ is purely $\phi_2$ type around here, and therefore, 
does not couple to the charged leptons and bottom quarks.
Furthermore, it should also be recalled that a common feature to all graphs is that the $H_1$ is mostly CP-mixed for small $M_{H^\pm}$ values but after the cross over around $M_{H^\pm}=145$ GeV it is mostly CP-even (see Fig. 5 of Ref.~\cite{Hesselbach:2007en}), this also impinging on the shape of the overall cross section. Finally, while in many cases the entire allowed range of the cross section can be spanned by simply considering both signs of $\mu$ (the two CP-conserving limits, $\phi_\mu=0$ and $180^{\rm o}$), where the results for CP-violating scenarios lie between these two cases, in many of the instances studied, owing to the structure of the CP-violating couplings between squark and Higgs states (which involve two terms with, at times, contrasting signs), this is generally not the case when $M_{H^\pm}<145$ GeV and also in some instances when $M_{H^\pm}> 145$ GeV (see, e.g., the right hand side of Fig.~6).  

\begin{figure}[h] \vskip 8cm \includegraphics{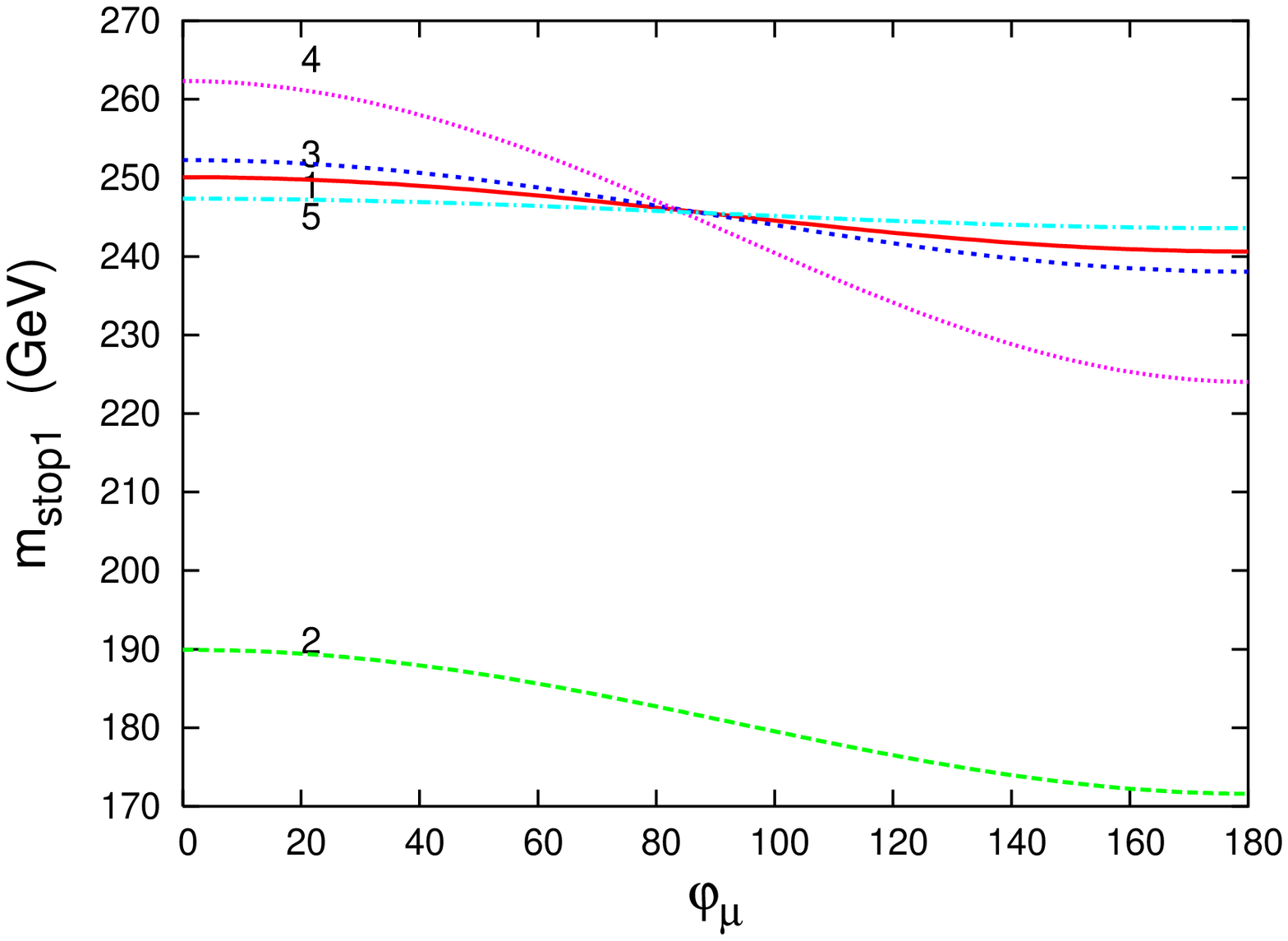} \caption{Variation of $m_{\tilde t_1}$ with $\phi_\mu$. Case 1: $\tan\beta=20$, $\mu=1$ TeV, $A_f=1$ TeV.  Case 2: $\tan\beta=20$, $\mu=1$ TeV, $A_f=1.5$ TeV.  Case 3: $\tan\beta=20$, $\mu=1.5$ TeV, $A_f=1$ TeV.  Case 4: $\tan\beta=5$, $\mu=1$ TeV, $A_f=1$ TeV. Case 5: $\tan\beta=50$, $\mu=1$ TeV, $A_f=1$ TeV.  In all cases $M_{U_3}=250$ GeV.} \label{fig:stopmass} \end{figure}

In addition to the stop mass dependence of the $gg\to H_1\to\gamma\gamma$ cross section, we have also studied how the latter varies with the masses of the other (s)particles entering the loops. However, in line with the results of  Refs.~\cite{Moretti:2007th}--\cite{Hesselbach:2007gf} for the case of the $H_1\to\gamma\gamma$ decay, we have found that their impact is largely negligible here too, no matter the value of the CP-violating phases. Nonetheless, one last aspect ought to be explored in connection with the $m_{\tilde t_1}$ dependence: how the value of the latter is itself affected by the CP-violating phases. In Fig.~\ref{fig:stopmass} we plot the variation of the lightest stop  mass with the  phase of $\mu$ for different parameter space points considered in the  earlier plots. For example, if $\tan\beta=20$, when $\mu=1.5$ TeV and  $A_f=1$ TeV, the mass $m_{\tilde t_1}$ lies within $\sim\pm 2\%$  of its average value whereas, for $\mu=1$ TeV and $A_f=1.5$ TeV, it lies within $\sim\pm 5\%$ of it. For $\tan\beta=50$ the spread is even narrower  ($\sim\pm 1\%$) while for $\tan\beta=5$ it is larger ($\sim\pm 10\%$). Notice though that these variations are typically smaller than the experimental resolution expected at the LHC in measuring $m_{\tilde t_1}$, which are of order 20\% or so  \cite{ATLAS,CMS}. Therefore, an intriguing prospect appears, that  the discovery of a light MSSM Higgs boson (with mass below 130 GeV or so) at the LHC may eventually enable one to disentangle the CP-violating case from the CP-conserving one,  so long that the relevant SUSY parameters entering  $gg\to H_1 \to\gamma\gamma$ are measured in the same process or elsewhere, in particular $M_{H_1}$ (or alternatively $M_{H^\pm}$, recall our Fig. 2 here and/or see Ref.~\cite{Hesselbach:2007en}  for the relation between the two) and $m_{\tilde t_1}$. This is not phenomenologically inconceivable, as, on the one hand, the   $\gamma\gamma$ final state is ideally suited to measure the Higgs mass (or indeed resolve superpositions of Higgs states) given the high di-photon mass resolution, and, on the other hand,  this Higgs detection mode requires a 
very high luminosity, unlike the discovery of those sparticles  (and the measurement of their masses and couplings) that 
enter the loops, chiefly top squarks.    

Finally, notice that it is from this point of view that we
decided to neglect studying here the impact of both reducible and irreducible backgrounds onto our signal process.
(The former typically include $\gamma$~jet and di-jet production, with jets mis-tagged as photons, whereas
the latter involve the tree-level $q\bar q\to\gamma\gamma$ and one-loop $gg\to\gamma\gamma$ [via box diagrams
involving quarks and squarks] subprocesses.) 
On the one hand, any CP-violating effect proportional to squark flavour 
mixing mass terms coming from non-resonant diagrams, such as the 
gluon-gluon-gamma-gamma box diagram, is expected to be very small owing to 
the stringent experimental constraints on rare CP and Flavour Changing Neutral Current (FCNC) processes.
On the other hand, since the background processes including CP-violating effects at higher order do not produce any kinematic structure
around the presumed $H_1$ resonance, they can accurately be measured off peak and smoothly interpolated beneath it.

\section{Conclusions}  

We have shown the conspicuous effects of CP violation onto the LHC cross section for $gg\to H_1\to \gamma\gamma$ in the MSSM, as the overall production and decay rate varies over an order or so of magnitude, depending on the value of $\phi_\mu$ (or, alternatively, $\phi_A$). Varying the mass of  the top squark from about 1 TeV to around 250 GeV has also a strong quantitative impact on  the cross section, possibly changing its dependence on the phases also  qualitatively. In fact, a significant dependence of the overall cross section on $\phi_\mu$ when $H_1$ is not a CP-mixed state (as seen from all the figures presented for large $M_{H^\pm}$  values) clearly exemplifies the importance of the Higgs-stop-stop coupling when CP violation is considered. Altogether then, it is clear that the prime detection channel of a light MSSM Higgs boson at the LHC is also a prime candidate to prove the existence of explicit CP violation in  minimal SUSY. In fact, the $\phi_\mu$ dependence of our results is strongest when the cross section is away from the minima, typically for $M_{H^\pm}$ larger than 150 GeV or so.   
\section{Acknowledgments}  

S. Munir's research is sponsored by DGAPA, UNAM, and   CONACyT, Mexico, project no. 82291-F.  

\vskip 5mm
\noindent
{\large \bf Appendix: Higgs couplings}\\[3mm]

Some of the Higgs couplings relevant to the process $gg\rightarrow H\rightarrow \gamma\gamma$ are given below.
For a full list of the couplings we refer to \cite{Lee:2007gn}.

\vskip 3mm
\noindent
\underline{Higgs-fermion-fermion couplings}\\[2mm]
\begin{equation}
{\cal L}_{H_i\bar f f}=-\sum_{f=u,d,l}~\frac{g~m_f}{2M_W}\sum_{i=1}^3H_i\bar f\left(
g^S_{H_i\bar f f}+ig^P_{H_i\bar f f}\gamma^5\right)f,
\label{eqn:Hiff}
\end{equation}
where $g=e/\sin\theta_W$ is the weak gauge coupling, 
$(g^S,g^P)=\left(O_{\phi_1i}/\cos\beta,-O_{ai}\tan\beta\right)$ for $f=d,l$ and 
$(g^S,g^P)=\left(O_{\phi_2i}/\sin\beta,-O_{ai}\cot\beta \right)$ for $f=u$. Here $u$ corresponds to 
$up$, $charm$, $top$ quarks, $d$ corresponds to $down$, $strange$, $bottom$ quarks and
$l$ stands for $e^-$, $\mu^-$, $\tau^-$ leptons.
The matrix $O$ is a real $3\times 3$ matrix which takes the Higgs fields from their 
electroweak eigenstates to the physical mass eigenstates:
\begin{equation}
\left(\phi_1,\phi_2,a\right)^T=O~\left(H_1,H_2,H_3\right)^T.
\end{equation}
\vskip 3mm
\noindent
\underline{Higgs-sfermion-sfermion couplings}\\[2mm]
\begin{equation}
{\cal L}_{H_i\tilde f \tilde f}=v\sum_{f=u,d}~g_{H_i\tilde f^*_j \tilde f_k}~~H_i\tilde f^*_k\tilde f_k.
\label{eqn:Hisfsf}
\end{equation}

Here \[vg_{H_i\tilde f^*_j \tilde f_k}=\left(\Gamma ^{\alpha \tilde f^*\tilde f}\right)_{\beta\gamma}
O_{\alpha i}U^{\tilde f^*}_{\beta j}U^{\tilde f}_{\gamma k},\] 
where $\alpha, i=1,2,3$; $j,k=1,2$ and $\beta,\gamma=L,R$. The matrix $U$ relates the gauge eigenstates
of the sfermions to their mass eigenstates:
\begin{equation} 
\left(\tilde f_L, \tilde f_R\right)^T_\alpha = U^{\tilde f}_{\alpha j}\left(\tilde f_1,\tilde f_2\right)^T_j.
\end{equation}

The couplings $\Gamma^{\alpha \tilde t^*\tilde t}$ in the $(\tilde t_L,\tilde t_R)$ basis are given by
\begin{eqnarray}
\Gamma^{a \tilde t^*\tilde t}&=&\frac{1}{\sqrt{2}}\left(
\begin{array}{cc}
0&ih^*_t~(\cos\beta~ A_t^*+\sin\beta~\mu)\\
-ih^*_t~(\cos\beta~ A_t+\sin\beta~\mu^*)&0 
\end{array}\right), \nonumber\\
\Gamma^{\phi_1 \tilde t^*\tilde t}&=&\left(
\begin{array}{cc}
-\frac{1}{4}\left(g^2-\frac{1}{3}g'^2\right)v\cos\beta&\frac{1}{\sqrt{2}}h^*_t\mu\\
\frac{1}{\sqrt{2}}h_t\mu^* & -\frac{1}{3}g'^2v\cos\beta
\end{array}\right), \nonumber\\
\Gamma^{\phi_2 \tilde t^*\tilde t}&=&\left(
\begin{array}{cc}
\left[-|h_t|^2+\frac{1}{4}\left(g^2-\frac{1}{3}g'^2\right)\right]v\sin\beta&-\frac{1}{\sqrt{2}}h^*_tA_t^*\\
-\frac{1}{\sqrt{2}}h_tA_t & \left[-|h_t|^2+\frac{1}{3}g'^2\right]v\sin\beta
\end{array}\right).
\end{eqnarray}

Here $h_t$ is the top Yukawa coupling, $g=e/\sin\theta_W$ and $g'=e/\cos\theta_W$.
Expressions for the couplings $\Gamma^{\alpha \tilde f^*\tilde f}$ in the case of other third
generation sfermions (which are of relevance to our study) are given in, for example, \cite{Lee:2007gn}.

  \end{document}